# The Effects of Dynamic Learning and the Forgetting Process on an Optimizing Modelling for Full-Service Repair Pricing Contracts for Medical Devices

Aiping Jiang[a]*, Lin Li[b], Xuemin Xu[a] and David Y.C. Huang[c]

[a]*SILC Business School, Shanghai University, Shanghai, China;*

[b]*Center for Human Development, University of California, San Diego, La Jolla, CA, USA;*

[c]*Medical Physics Graduate Program, Duke Kunshan University, Kunshan, China*

E-mail of the *corresponding author: ap724@shu.edu.cn

# The Effects of Dynamic Learning and the Forgetting Process on an Optimizing Modelling for Full-Service Repair Pricing Contracts for Medical Devices

In order to improve the profitability and customer service management of original equipment manufacturers (OEMs) in a market where full-service (FS) and on-call service (OS) co-exist, this article extends the optimizing modelling for pricing FS repair contracts with the effects of dynamic learning and forgetting. Along with considering autonomous learning in maintenance practice, this study also analyses how induced learning and forgetting process in a workplace put impact on the pricing optimizing model of FS contracts in the portfolio of FS and OS. A numerical analysis based on real data from a medical industry proves that the enhanced FS pricing model discussed here has two main advantages: (1) It could prominently improve repair efficiency, and (2) It help OEMs gain better profits compared to the original FS model and the sole OS maintenance. Sensitivity analysis shows that if internal failure rate increases, the optimized FS price rises gradually until reaching the maximum value, and profitability to the OEM increases overall; if frequency of induced learning goes up, the optimal FS price rises after a short-term downward trend, with a stable profitability to the OEM.

Keywords: original equipment manufacturers (OEMs); full-service (FS) maintenance; learning and forgetting; optimization; pricing

## 1. Introduction

Large medical devices purchased by medical institutions (e.g. CT, MRI machines and medical linear accelerators are expensive and usually incur significant maintenance cost, a large amount of which comes from the service interruption due to irregular breakdown, and unpredictable repairs time (ECRI Institute, 2013). Therefore, the medical institutes, for example tertiary hospitals, are eager to lower the uncertainty of maintenance so that the devices can have more stable performance and longer service life. Canek and Rodrigo (2021) claimed that although having internal maintenance team can help reduce uncertainty, it is often uneconomical for the equipment operator to have such specialist tools and personnel in-house. In those instances, maintenance operations are carried out by the Original Equipment Manufacturer (OEM) or another non-OEM service provider. In response to the market demands, Laurens et. al (2021) found that more and more OEMs shifts their focus from solely selling products to products bundled with end maintenance solutions. But the idea is barely new, Huber and Spinler (2012) have identified a portfolio of FS and OS maintenance contracts. The selection between those two forms of contracts depends on clients' risk appetite.

Clients with low risk aversion are likely to sign OS contracts because they could tolerate the risk of catastrophic cost of large failure in exchange of no investment costly and inflexible prepaid maintenance contract when purchasing the devices; whereas clients with high risk-aversion tend to sign FS contracts because they are unwilling to take risks of potentially large amount of contingent maintenance cost. From the perspective of OEMs, Noble and Gruca (1999) first proposed a cost-plus-profit margin pricing method for OS contracts, which is easy and simple to manage . However, there are inevitable drawbacks with OS contracts too. First, the maintenance schedules of OS are usually inconsistent, bringing the OEMs a more difficult time to plan and prepare for each service so that leading to inferior work quality and efficiency (Roels et al., 2010). In addition, Müller and Turner (2005) pointed out there is little motivation or will for OEMs to reduce the cost under OS contracts because the maintenance price is equal to total cost plus fixed profit margin.

Huber and Spinler (2012; 2014) found that there is an ongoing trend for the research in industrial FS contracts. FS contracts relieve customers from uncertain maintenance costs through covering all maintenance related costs during a predetermined horizon in exchange for a fixed service fee Laurens et. al (2021). At the same time, Deprez et al (2021) studied FS contracts from the viewpoint of the service provider, i.e. OEMs.

Under a FS contract OEMs shall have full understanding of their own products and are able to evaluate the operational environment when setting devices up. The information is helpful for them to predict costs and even arrange preventive maintenance before failure happens. However, FS contracts also bring OEMs a few challenges. Kim et al. (2007) claimed that failure rate and repair costs are usually difficult to predict, and the FS contracts totally transfer the risk of uncertainty which was previously undertaken by clients to OEMs. Therefore, OEMs have strong motivation to control uncertainty and improve efficiency to maximize profitability based on fixed price under FS contracts.

Inspired by the challenges of FS discussed above, some OEMs began to adopt the idea of workplace learning which could increase the working efficiency of their technicians. Wang and Lee (2001) initially analysed an effect learning curve to measure learning rates for total productive maintenance operations. Then, the scope of research on learning effects in industries is continuously expanding. Application of learning effects are examined by scholars in different research fields, including production scheme (Glock et al., 2012), inventory control (Jaber & Glock, 2013; Teng et al., 2014), and quality enhancement (Lolli et al., 2016). If firms have a sharper learning curve, they may be more competitive in market in the long run (Biskup & Simons, 2004). Tarakci et al. (2009) stated that learning by practice contribute positively to OEMs' profitability; Lolli et al. (2016) also agreed that autonomous learning could benefit both productivity and quality of work. However, some scholars also acknowledged the limits of autonomous learning (Nasr and Jaber, 2019). Asadayoobi, N., et al. (2020) and Fu, K, et al (2020) found that fatigue and forgetting effects exist over the autonomous learning process; in management terminology, there are diminishing marginal effects towards autonomous learning. . Research also showed that the technicians' learning efficiency directly contributes to the success of their firms in competitive markets. (Peltokorpi and Jaber 2020) , thus how to prevent marginal effects diminishing or to compensate the loss become a hot topic for management. To address this, the concept of "dynamic learning" came to the world of industrial engineering and management. The dynamic learning process consists of two parts -- autonomous learning in day-to-day work and induced learning from the training provided out of the workplace.

Regarding the effects of dynamic learning on a mixed portfolio of OS and FS, Huber and Spinler (2014) presented that over time, maintenance technicians obtain opportunities to learn more cost-efficient repair techniques and employ cost-optimized maintenance schemes through accumulating job proficiency, as well as gaining better knowledge with overall working conditions and breakdown patterns of the equipment. Although their research clearly explained and robustly modelled how autonomous learning could be involved in the OEM's decision making process upon maintenance service, neither the effects of induced learning nor forgetting phenomena was covered in their studies.

This paper aims to fills this gap. This study prioritizes the topic of how OEMs maximize their profit through both autonomous learning discussed above, and the induced learning such as off-work training for technicians carrying out FS maintenance . Additionally, the paper also considers the effect of forgetting on workplace training due to interruptions, which is essential but haa seldom been discussed in maintenance procedures before. Based on data gathered from a public medical hospital in mainland China and a computer simulation, the authors raise an enhanced optimizing model, where profit is set as the target function with the decision variable FS price and other important constraints conditions. By comparing and contrasting the new model with the existing one (Huber & Spinler, 2012), the author justified the new model with a concept of dynamic learning and forgetting could improve the profitability by helping OEMs make better management strategies upon investment in external training and FS contracts pricing . Sensitivity analysis is also performed on two key parameters, internal failure and training frequency.

The remainder of this paper is organised as follows. Section 2 describes the model of maximizing OEMs' profits considering the effect of dynamic learning and forgetting. Section 3 presents the numerical analysis based on this model and section 4 makes a conclusion.

## 2. Model Formulation

First-line technicians and scholars point out that on-call services usually have a higher potential risk of instability, hence the persistent demand for less costly full-service contracts. Meanwhile, the high maintenance costs of medical equipment such as CT, MRI and radiotherapy devices is deemed a wicked problem by the medical institute. For OEMs, it has been proven that the OEMs completely offering FS repairs can benefit from working proficiency and by repetitively providing the repair and maintenance operation to clients. The OEMs thus have an opportunity to learn more cost-efficient repair techniques and employ cost-optimized maintenance schemes to finally earn higher profits with lower prices for clients.

In this model, the individual repair costs are stochastically independent of the number of failures. Stochastic repair costs over contract period $C_R$, a stochastic process, is likely to vary over the contract period due to the occurrence of different breakdowns. Equipment experiences a variety of failures caused by internal breakdowns of devices with different probabilities over stages of the contract period. The engineers of OEM companies are exposed to moral hazards and thus will probably delay the FS provided to clients. The clients, usually the machine users, face two service options: they may either sign a FS contract which stipulates a fixed price $P^{FS}$, or individually pay for each repair and maintenance activity through traditional on-call service. The selection depends on minimizing disutility rather than on solely considering price. Notation is summarized in Table 1.

**Table 1**. Notation used in the model

### *2.1 The costs of the OEM*

#### *2.1.1 Cost of repair and maintenance*

To focus on repair and maintenance operation expenses, the OEM incurs individual repair costs $C_r$, and Huber and Spinler (2012) created the equation for repair costs as:

$$E\left[\widetilde{C}_R(B)\right] = \sum_{j=1}^{Z} E\left[\widetilde{C}_{rj}\right] \cdot E\left[\widetilde{N}_j(M)\right] \tag{1}$$

Huber and Spinler (2014) extended this equation to match a multi-period model with $Z$ consecutive intervals of time length $t_j$. The number of expected failures for the $j$ th interval is given by:

$$E\left[\widetilde{N}_j(M)\right] = \Phi_{j-1}^{\text{int}} t_j + \frac{t_j^2 g_j}{2}\left[(1-\rho) + \frac{\rho}{M}\right]$$

$$= \Phi_0^{\text{int}} t_j + \frac{\rho t_j^2 g_j}{2M} + \frac{t_j^2 (1-\rho)}{2}\left(g_j + 2 \cdot \sum_{k=0}^{j-1} g_k\right) \tag{2}$$

$\rho$ (0< $\rho$ <1) is a factor measuring improvement gained through preventive maintenance; $\rho$ = 0 refers to a state after imperfect maintenance meaning as “bad as old”, while $\rho$ = 1 indicates perfect maintenance referred to as “good as new”.

Based on the Equations (1) and (2) raised by Huber and Spinler (2014), service providers can optimize the number of preventive maintenance (PM) activities (noted as $M$ ), in the following equation:

$$M^{*} = \sqrt{\frac{\rho t_j^{M^2}}{2\bar{c}_m Z} \sum_{j=1}^{Z} E[\tilde{C}_{rj}] g_i} \quad (3)$$

where $t_j^M$ is the length of one maintenance period $j$ , which is different from $t_j$ (the length of period $j$ ), and a day is counted as 24 hours. The optimal PM activities $M$ is given in the nearest integer.

Although Huber and Spinler (2014) demonstrated that the aging factor $g_i$ may not be constant at different running stages of devices, there lacks specific information about what the distribution of the rate should be, which may lead the setting of these essential parameters to be arbitrary. In this paper, because the clients usually make the decision whether to sign a FS contract when the equipment is newly installed, there exists a run-in period by which the failure rate $\Phi_j^{\text{int}}$ should be shaped as a bathtub curve (Lienig & Bruemmer, 2017). $g_i$ can be a piece-wise equation of the Hazard Ratio $h_{(t)}$ of the Weibull distribution.

To formulate the three-stage failure rate with the shape of a bathtub curve demonstrated above, the aging factor $g_i$ of the failure rate should change along with time. Although device operators are becoming increasingly familiar with equipment operation during the run-in periods, the marginal returns of learning from work diminish (Munford & Shahani, 1973). Therefore, in the first stage, slope $g_i$ is negative but with decreasing absolute values. However, in the second stage, the failure rate often becomes constant that $g_i$ should equal 0. As for the third stage which is opposite to the first, the failure rate rises sharply with an increasing marginal effect. Hence, $g_i$ can be formulated as a piece-wise equation of the Hazard Ratio $h_{(t)}$ of the Weibull distribution demonstrated in Weibull (1951):

$$g_j = \begin{cases} \frac{-k_1}{m}\left(\frac{j-0}{m}\right)^{k_1-1} & (1 \le j \le z_1) \quad 0 < k_1 < 1 \quad t_0 > 0 \\ 0 & (z_1 < j \le z_2) \\ \frac{k_2}{m}\left(\frac{j-z_2}{m}\right)^{k_2-1} & (z_2 \le j \le z_3) \quad 0 < k_2 < 1 \quad t_0 > 0 \end{cases} \tag{4}$$

The shape parameter is $g_i$ but with an adjustment on sign, $m$ is the scale parameter and $j$ is the period index. $g_i$ is also exactly the slope of equation $\phi_j^{\text{int}}(t) = \phi_{(j-1)}^{\text{int}} + g_j \cdot t_j \quad (j \ge 1)$. In this way, the total expected FS repair cost with optimal PM activity is:

$$E\left[\widetilde{C}_R(B)\right] = \sum_{j=1}^{z} E\left[\widetilde{C}_{rj}\right] \cdot E\left[\widetilde{N}_j\left(M^*\right)\right] \tag{5}$$

Here $E\left[\widetilde{N}_j\left(M^*\right)\right]$ is the expected failure time in the $j$ th interval under optimal numbers of PM, and $E\left[\widetilde{C}_{rj}\right]$ is the expected individual repair cost of the $j$ th interval. To simplify calculation, this paper assumes that $E\left[\widetilde{C}_R(B)\right] = E\left[\widetilde{C}_R^{OS}\right]$. Moreover, optimal preventive maintenance costs can be depicted as $\widetilde{C}_M^{FS}\left(M^*\right) = \bar{c}_M \cdot (M^* - 1)$. $\bar{c}_M$ refers to average maintenance cost per single maintenance.

*2.1.2 Cost of delay*

The engineers of OEM company are exposed to moral hazards as they have an incentive to prioritize OS and thus delay FS clients. To address this issue, a repair time guarantee is introduced where any repair time in excess of the contracted repair time is reimbursed to the customer. The resulting expected delay cost is given as:

$$E\left[\widetilde{C}_D^{FS}(B)\right] = \sum_{j=1}^{z} E\left[\widetilde{N}^{FS}(B)\right] \cdot E\left[\widetilde{C}_d\right] \quad (6)$$

Here, $E\left[\widetilde{N}^{FS}(B)\right] = \sum_{j=1}^{z} E\left[\widetilde{N}_j(M)\right]$

While the second term reflects individual delay costs which used to be modelled in terms of the queuing approach adopted in Huber and Spinler (2012), a shortage of repair technicians may be incurred due to insufficient professional human resources. However, the devices studied in this article have relatively low average failure rate, and most run no more than 8 hours per day according to staff's maximum working time. In addition, the OEM company in this paper is considered as a significantly large player in the medical devices manufacturing industry, so in general, enough practitioners are available to provide services in time. Due to this, delays could be a stochastic process closely influenced by external factors with very low probability, such as extreme weather, traffic congestion or accident, workers strike. There is a provision that the tolerance rate of inevitable delays further protects the FS provider from extra costs derived from too heavy delay penalties, which helps maintain their motivation to provide full-service repairs. Another provision is that none of the actions taken by OEM companies or clients will affect the individual delay cost $C_d$.

### *2.2 Modelling of extended learning and forgetting effects*

In the section 2.1, we have formulated the basic repair cost equation as the sum of the product of the expected failure number $E\left[\widetilde{N}_j\right]$ and the unit repair cost $E\left[\widetilde{C}_{rj}\right]$. Some scholars point out that the effects of learning may affect both time and cost of repairs.

Huber and Spinler (2014) raised the idea that FS repair efficiency could be improved through learning. They also demonstrated that the learning effect on costs is a function of the operation time of equipment, rather than the number of repairs or maintenance. Huber and Spinler (2014) came up with the learning effect formula by referring to Wright (1936), defined as $j^{-\alpha}$, where $j$ is the period index $j \in \{1, 2, \cdots, z\}$. $\alpha$ is the learning factor. With the introduction of learning effects that only exist in full-service repairs, the repair cost equation can be specifically adjusted for the FS contract, as below:

$$E\left[\widetilde{C}_R^{FS}(B)\right] = E\left[\widetilde{C}_R(B)\right] \cdot A(B) = \sum_{j=1}^{z} E\left[\widetilde{N}_j\right] \cdot E\left[\widetilde{C}_{rj}\right] \cdot A(B) \tag{7}$$

where $A(B) < 1$ represents the cumulative learning effect over the equipment lifetime. From the combined optimal maintenance schedule and the equation deduced in the previous section, the repair cost equation adjusted for FS was expanded by Huber and Spinler (2014) as:

$$\begin{aligned} E\left[\widetilde{C}_R^{FS}(B, M^*)\right] &= E\left[\widetilde{C}_R(B, M^*)\right] \cdot A(B) \\ &= E\left[\widetilde{C}_R(B, M^*)\right] \cdot j^{-\alpha} \end{aligned} \tag{8}$$

This is often referred to as autonomous learning, or "learning-by-doing", since improvement is a by-product of ongoing production.

However, there are obvious limitations in Huber and Spinler's model of the learning effect. Firstly, it deals only with autonomous learning. The induced component is not considered despite its relevance as competitive leverage and it does not consider the forgetting component. Secondly, Huber and Spinler (2014) conservatively assumed that repair providers make no investment or provisions for training programs for their engineers or technicians. In other words, the learning effect described in the study by Huber and Spinler (2014) only refers to passive learning. Additionally, they gave no consideration to the unavoidable forgetting phenomena in realistic practices, like interruptions in the learning process due to the reasons such as unexpected breakdowns or PMs that would also influence FS repair efficiency (Cui et al., 2018).

To compensate the drawbacks, the simplified learning effect function that consists only of passive learning depending on time $j^{-\alpha}$ , can be dichotomized into binary variables of autonomous learning gained by on-site repairs or maintenance operations $N(t)$, and induced learning by training $L(t)$, respectively. This paper assumes that ( $N(t)$ and $L(t)$) are both proportional functions of time $t$, and it introduces a learning exponent for each approach, extending the original passive learning factor $\alpha$ , as $\alpha_{Auto}$ or $\alpha_{indu}$. The authors hold the assumption that the learning effect function complies with the negative exponential distribution, thus the learning effect equation for FS contract becomes:

$$A(B)=\left[N(t)\right]^{-\alpha}\left[L(t)\right]^{-\alpha_{indu}} \tag{9}$$

The simplified autonomous learning formula $j^{-\alpha}$ is replaced with the newly inferred $N(t)$. Unlike the original function whose independent variable is exactly the period index $j$, the variable in the extended learning effect function $N(t)$ is equivalent to repair time $t^r$, and $L(t)$ is equivalent to the total effective training time calculated as gross training time $t^l$ minus equivalent forgetting time $t^f$. The existence of forgetting effects refers to the imperfect substitute of induced learning as forgetting is a natural phenomenon in all kinds of external gains. Then the equation below introduces all factors the authors plan to analyze, including autonomous learning, induced learning and forgetting effect:

$$A(B) = \left[ N\left(t^r\right) \right]^{-\alpha_{Auto}} \left[ L\left(t^l - t^f\right) \right]^{-\alpha_{indu}} = \left(t^r\right)^{-\alpha_{Auto}} \cdot \left(t^l - t^f\right)^{-\alpha_{indu}} \tag{10}$$

The equation below illustrates how total repair time and total effective training time are calculated. $t_j^r$ refers to repair time in the $j$ th interval which equals to $\Phi_j^{\text{int}} \cdot t_j$ and the total repair time is simply the sum of $j, j \in \{1, 2, ..., z\}$ repair costs.

$$t^r = \sum_{j=1}^{z} t_j^r = \sum_{j=1}^{z} \Phi_j^{\text{int}} \cdot t_j \tag{11}$$

However, the formulation of total effective training time would be more complex. Training time in the $j$ th interval must be no longer than the length of that interval minus time for maintenance and repair, resulting from both the internal and external problems, which is defined as "surplus time". The equation is illustrated as:

$$t_j^l \le t_j - t_j^M - \left(\Phi_j^{\text{int}} \cdot t_j + \Phi_j^{\text{int}} \cdot t_j\right) \tag{12}$$

Meanwhile, the $lf$ should not be too large as the training time usually occupies the working hours, or it should not be too small to ensure the effectiveness and quality. Thus, another variable depicting the proportion of surplus time for the training, as "$lf$", $0 < lf < 1$, which will be named as the "learning rate" in the equation below. $lf$ might vary from company to company or from time to time, but the value should be in a reasonable range, as explained above. Hence, the authors form the training time in $j$ th interval as:

$$t_j^l = \left[ t_j - \left(\Phi_j^{ext} \cdot t_j + \Phi_j^{\text{int}} \cdot t_j + t_j^M\right) \right] \cdot lf \tag{13}$$

Total training time equals:

$$t^l = \left[ \sum_{j=1}^{z} t_j - \sum_{j=1}^{z} \left( \Phi_j^{ext} \cdot t_j + \Phi_j^{\text{int}} \cdot t_j + t_j^M \right) \right] lf \tag{14}$$

There have already been several studies on modelling the forgetting effects. One of the first attempts was made by Carlson and Rowe (1976), and they raised a formula similar to Wright's learning curve as $x^f$. This approach was validated some years later by Globerson et al. (1989), but one obvious drawback is that the likelihood of failures is not considered. To remedy this, the quality-based element in the learning curves was firstly introduced by Jaber and Guiffrida (2004), who extended Wright's law with the hypothesis that there is likely to be failure of equipment which interrupts normal production, and such interruptions will inevitably lead to extra costs. However, the previous studies are not complete because of the complex to model the forgetting effects. In this paper, the failure of equipment requiring immediate repairs can be regarded as an interruption of the training process. The uncontrollable inconsistencies of training should be removed from total training time as the Equation (14) shows, and its negative impact on training time is defined as the forgetting effect in this case. However, Jaber and Guiffrida (2004) do not distinguish between internal and external failures, and thus different failures attach different influences. The authors notice this weak point and divide the likelihood of failures into two parts, marked as $\Phi_j^{ext}$ and $\Phi_j^{\text{int}}$ in previous section:

$$t_j^f = \Phi_j^{ext} \cdot t_j + \Phi_j^{\text{int}} \cdot t_j + t_j^M \tag{15}$$

The equations in the previous research above hold the assumption that technicians cannot enhance knowledge or techniques gained from workplace training during on-site repair or maintenance. However, it doesn't apply to the industry studied in this paper. Considering the fact that engineers can practice the latest repair techniques learned from training in workplace to remedy the forgetting effect partially, the previous formulas about "learning by doing" can be adapted, which is expanded by Jaber and Guiffrida (2004) . Therefore, the following equation is suggested.

$$t_j^f = \Phi_j^{ext} \cdot t_j + 2 \cdot \left( \frac{\Phi_j^{\text{int}}}{2} \right)^{1-\varepsilon} \cdot \left[ \left( t_j - \Phi_j^{ext} \cdot t_j \right) \cdot lf \right]^{1-2\varepsilon} \tag{16}$$

where ε is the learning exponent for putting knowledge learned from workplace training into practice of on-site repair work. In general, this exponent should be relatively smaller than induced learning factor because engineers apply incompletely what they learned in theory to their practice, regarded as "imperfect revision". $\left( t_j - \phi_j^{ext} \cdot t_j \right) \cdot lf$ in equation (16) refers to the training time on the ideal condition that there were no repair or maintenance works, in the place of x in the original formula formulated by Carlson and Rowe (1976). However, external failures such as strikes, protests of engineers and extreme conditions will not provide remedies for forgetting effects. Therefore, the total forgetting effect on training time can be:

$$t^f = \sum_{j=1}^{z} \left[ \Phi_j^{ext} \cdot t_j + 2 \cdot \left( \frac{\Phi_j^{\text{int}}}{2} \right)^{1-\varepsilon} \cdot \left[ \left( t_j - \Phi_j^{ext} \cdot t_j \right) \cdot lf \right]^{1-2\varepsilon} \right] \tag{17}$$

An industry report issued by the American Society for Training and Development (ASTD) indicates that U.S. organizations have an annual expenditure of $126 billion on employee training and development (Paradise & Goswami, 2007). Thoughtful investments in induced learning activities can augment the overall learning rate in practice, thereby accelerating the benefits by continuous process improvement. Then, total costs depend on induced learning units via a linear cost structure. For OEM companies, the total investment in training during the planning period $(t-1, t)$ is:

$$C^l = \sum_{j=1}^{z} l\left[L(j) - L(j-1)\right] = \sum_{j=1}^{z} l \cdot t_j^l \qquad (18)$$

where $l$ is the training cost coefficient, which indicates unit training costs and can be estimated using historical cost data.

### *2.3 Improving FS efficiency through optimized training frequency*

In the last subsection, the authors have taken both the theoretical and functional derivations of the learning and forgetting effects for improving the efficiency of FS repair in detail. The training frequency $lf$ has been introduced as an essential variable to represent the proportion of engineers' time that can be available for training activities.

Based on the mathematical deductions given in the sections above, the total FS costs can be formulated as $\mathrm{E}\left[\tilde{\mathrm{C}}_{\mathrm{R}}(\mathrm{B},\mathrm{M}^*)\right] \cdot A(B) + E\left[\tilde{C}_D^{FS}(M^*)\right] + \tilde{C}_M^{FS}(M^*) + C^l$. Holding all other factors constant in their optimal values, total FS cost can be regarded as a function of learning effect $A(B)$ and total learning $C^l$. According to Equations (10), (14) and (17), $lf$ has a non-linear correlation with the cumulative learning effect $A(B)$. Specifically, $A(B)$ is a power function of $lf$ with a negative exponent $-\alpha_{indu}$, 。 However, according to Equations (13) and (14), the total training cost $C^l$ and learning rate $lf$ are proportionally correlated and $lf$ is an independent variable. The combination of both linear and non-linear relationships means a more precise way to directly judge if an optimal learning rate exists which could minimize repair costs for the OEM to create maximum profit if prices remain fixed.

**Corollary 1:**

The expected total FS cost which is represented as $E\left[\tilde{C}_R^{FS}\left(B,M^*\right)\right] = \mathrm{E}\left[\tilde{\mathrm{C}}_{\mathrm{R}}(\mathrm{B},M^*)\right]\cdot A(B) + E\left[\tilde{C}_D^{FS}(M^*)\right] + \tilde{C}_M^{FS}(M^*) + C^l$ should begin with a downward trend until the lowest point is reached, as the training frequency $lf$ increases when the value of $lf$ is relatively small. Once the training frequency $lf$ exceeds the turning points, the FS cost would rise. In other words, an optimal full-service contract cost to OEMs should exist.

**Proof of Corollary 1** :

First, given

$t^r = \sum_{j=1}^{z} t_j^r = \sum_{j=1}^{z} \Phi_j^{\mathrm{int}} \cdot t_j$,

$t^l = \left[\sum_{j=1}^{z} t_j - \sum_{j=1}^{z}\left(\Phi_j^{ext}\cdot t_j + \Phi_j^{\mathrm{int}}\cdot t_j + t_j^M\right)\right] lf$,

$t_j^f = \Phi_j^{ext}\cdot t_j + \Phi_j^{\mathrm{int}}\cdot t_j + t_j^M$,

the cumulative learning effect equation $A(B) = \left(t^r\right)\frac{-\alpha}{auto}.\left(t^l - t^f\right)\frac{-\alpha}{indu}$ can be expanded as:

$$A(B) = \left(\sum_{j=1}^{z} \Phi_j^{\mathrm{int}}\cdot t_j\right)^{-\alpha_{Auto}} \cdot \begin{Bmatrix} \left[\sum_{j=1}^{z} t_j - \sum_{j=1}^{z}\left(\Phi_j^{ext}\cdot t_j + \Phi_j^{\mathrm{int}}\cdot t_j + t_j^M\right)\right] lf \\ -\sum_{j=1}^{z}\left[\Phi_j^{ext}\cdot t_j + 2\left(\frac{\Phi_j^{\mathrm{int}}}{2}\right)^{1-\varepsilon}\left[\left(t_j - \Phi_j^{ext}\cdot t_j\right) lf\right]^{1-2\varepsilon}\right] \end{Bmatrix}^{-\alpha_{indu}}$$

Then, combining the Equations (14) and (18) altogether, total cost incurred by training can be deduced as:

$$C^l = l\left[\sum_{j=1}^{z} t_j - \sum_{j=1}^{z}\left(\Phi_j^{ext}\cdot t_j + \Phi_j^{\mathrm{int}}\cdot t_j + t_j^M\right)\right] lf$$

For further inference, the authors simplify the two equations above by replacing complex elements with capital letters.

Hence, $A(B)$ can be rewritten as：

$$A(B) = Q^{-\alpha_{Auto}}\left[(R - S - Q - U)lf - S - V\cdot lf^{1-2\varepsilon}\right]^{-\alpha_{indu}}$$

Here,

$Q=(\sum_{j=1}^{z}\phi_j^{\text{int}}t_j)$, , $R=\sum_{j=1}^{z}t_j$ $S=\sum_{j=1}^{z}\phi_j^{ext}t_j$, $U=\sum_{j=1}^{z}t_j^{M}$, $V=2\sum_{j=1}^{z}(\frac{\phi_j^{\text{int}}}{2})^{1-\varepsilon}(t_j-\phi_j^{ext}t_j)^{1-2\varepsilon}$, and assuming $(R-S-Q-U)\gg$ S,

Similarly,

$$C^{l}=l(R-S-Q-U)\cdot lf$$

Therefore, the total expected FS cost can be adapted as:

$$E\left[\tilde{C}^{FS}\left(B,M^{*}\right)\right]=\mathrm{E}\left[\tilde{\mathrm{C}}_{\mathrm{R}}(\mathrm{B,M}^{*})\right]\cdot A(B)+E\left[\tilde{C}_{D}^{FS}(M^{*})\right]+\tilde{C}_{M}^{FS}(M^{*})+C^{l}$$

$$=E\left[\tilde{C}_{R}(B,M^{*})\right]\cdot Q^{-\alpha_{Auto}}\cdot\left[(R-S-Q-U)\cdot lf-S-V\cdot lf^{1-2\varepsilon}\right]^{-\alpha_{indu}}$$

$$+E\left[\tilde{C}_{D}^{FS}(M^{*})\right]+\tilde{C}_{M}^{FS}(M^{*})+l\cdot(R-S-Q-U)\cdot lf$$

The aim of this section is to optimize the $lf$ so that a minimum expected FS cost $E\left[\tilde{C}^{FS}\left(B,M^{*}\right)\right]$ would be reached,

which can be written as

$$\underset{lf\in(0,1)}{\arg\min}\,E\left[\tilde{C}^{FS}\left(B,M^{*}\right)\right]=\mathrm{E}\left[\tilde{C}_{\mathrm{R}}(\mathrm{B},M^{*})\right]\cdot Q^{-\alpha_{Auto}}\cdot\left[(R-S-Q-U)\cdot lf-S-V\cdot lf^{1-2\varepsilon}\right]^{-\alpha_{indu}}$$
$$+E\left[\tilde{C}_{D}^{FS}(M^{*})\right]+\tilde{C}_{M}^{FS}(M^{*})+l\cdot(R-S-Q-U)\cdot lf$$

Next, the authors begin with taking a derivative with respect to training frequency $lf$ as in order to obtain the monotonicity: *Applying the chain rule:*

$$\frac{\partial E\left[\tilde{C}^{FS}(B)\right]}{\partial lf}=E\left[\tilde{C}_{\mathrm{R}}(\mathrm{B},M^{*})\right]\cdot Q^{-\alpha_{Auto}}$$

$$\cdot(-\alpha_{Auto})\cdot\left[(R-S-Q-U)\cdot lf-S-V\cdot lf^{1-2\varepsilon}\right]^{(-\alpha_{indu}-1)}$$
$$\cdot\left[(R-S-Q-U)-(1-2\varepsilon)\cdot V\cdot lf^{-2\varepsilon}\right]+l\cdot(R-S-Q-U)$$

$$=\underline{l}\cdot\left(\underline{R-S-Q-U}\right)\underline{\pm}(\underline{-\alpha}_{\underline{Auto}})\underline{A}\left(\underline{B}\right)\underline{E}\left[\underline{\tilde{C}_{R}}(\underline{B},\underline{M^{*}})\right]\underline{\cdot}\frac{\left(\underline{R-S-Q-U}\right)-\left(\underline{1-2\varepsilon}\right)\cdot\underline{V}\cdot\underline{lf}^{\underline{-2\varepsilon}}}{\left(\underline{R-S-Q-U}\right)\cdot\underline{lf}-\underline{S-V\cdot lf}^{\underline{1-2\varepsilon}}}$$

$$=\underline{l}\cdot\left(\underline{R-S-Q-U}\right)\underline{\pm}(\underline{\alpha}_{\underline{Auto}})\underline{A}\left(\underline{B}\right)\underline{E}\left[\underline{\tilde{C}_{R}}(\underline{B},\underline{M^{*}})\right]\underline{\cdot}\frac{\left(\underline{1-2\varepsilon}\right)\cdot\underline{V}\cdot\underline{lf}^{\underline{-2\varepsilon}}\underline{-}\left(\underline{R-S-Q-U}\right)}{\left(\underline{R-S-Q-U}\right)\cdot\underline{lf}-\underline{S-V\cdot lf}^{\underline{1-2\varepsilon}}} \quad (eq\ a)$$

We assume that $R-S-Q-U\geq 0$ , $(1-2\varepsilon)\gg 0$ and $R-S-Q-U\gg S$ always hold,

$$(a)<\underline{l}\cdot\left(\underline{R-S-Q-U}\right)\underline{\pm}(\underline{\alpha}_{\underline{Auto}})\underline{A}\left(\underline{B}\right)\underline{E}\left[\underline{\tilde{C}_{R}}(\underline{B},\underline{M^{*}})\right]\underline{\cdot}\frac{\left(\underline{1-2\varepsilon}\right)\cdot\underline{V}\cdot\underline{lf}^{\underline{-2\varepsilon}}}{\left(\underline{R-S-Q-U}\right)\cdot\underline{lf}-\underline{V\cdot lf}^{\underline{1-2\varepsilon}}}$$

$$=\underline{l}\cdot\left(\underline{R-S-Q-U}\right)\underline{\pm}(\underline{\alpha}_{\underline{Auto}})\underline{A}\left(\underline{B}\right)\underline{E}\left[\underline{\tilde{C}_{R}}(\underline{B},\underline{M^{*}})\right]\underline{\cdot}\frac{\left(\underline{1-2\varepsilon}\right)\cdot\underline{V}}{\left(\underline{R-S-Q-U}\right)\cdot\underline{lf}^{\underline{1+2\varepsilon}}\underline{-V\cdot lf}}$$

$$=\underline{l}\cdot\left(\underline{R-S-Q-U}\right)\underline{\pm}(\underline{\alpha}_{\underline{Auto}})\underline{A}\left(\underline{B}\right)\underline{E}\left[\underline{\tilde{C}_{R}}(\underline{B},\underline{M^{*}})\right]\underline{\cdot}\frac{\left(\underline{1-2\varepsilon}\right)\cdot\underline{V}}{\left(\left(\underline{R-S-Q-U}\right)\cdot\underline{lf}^{\underline{2\varepsilon}}\underline{-V}\right)\cdot\underline{lf}}$$

$$\leq\underline{l}\cdot\left(\underline{R-S-Q-U}\right)\underline{\pm}(\underline{\alpha}_{\underline{Auto}})\underline{A}\left(\underline{B}\right)\underline{E}\left[\underline{\tilde{C}_{R}}(\underline{B},\underline{M^{*}})\right]\underline{\cdot}\frac{\left(\underline{1-2\varepsilon}\right)\cdot\underline{V}}{\left(\left(\underline{R-S-Q-U}\right)\cdot\underline{lf}-\underline{V}\right)\cdot\underline{lf}}(0<lf<1, and\, 0<\varepsilon\ll$$

According to the Vieta's formulas, the x-coordinate of the vertex for the quadratic equation $((R-S-Q-U)lf-V)lf\to 0$ is $lf\to\frac{V}{2(R-S-Q-U)}$

(Since $((R-S-Q-U)lf-V)lf$ is the determinator, it cannot be set as zero, $\rightarrow$ is applied here instead of "=" ). By this, there should exist a turning point of $lf$ around the neighbourhood of $\frac{V}{2(R-S-Q-U)}$ that lets the derivative $\frac{\partial\underline{E}\left[\underline{\tilde{C}^{FS}}(\underline{B})\right]}{\partial\underline{lf}}$ equals to 0.

Thus, a minimum expected FS cost $\underline{E}\left[\underline{\tilde{C}^{FS}}(\underline{B,M^{*}})\right]$ marked as $\underline{E*}\left[\underline{\tilde{C}^{FS}}(\underline{B,M^{*}})\right]$ is ultimately reached

2.4 Profit for the OEM

If clients minimize their disutility according to the equation raised by Kim et al. (2007), then the equation by Huber and Spinler (2014) can be applied as following:

$$\tilde{u}_i(B)=\begin{cases}(1+\beta)\left(E\left[\tilde{C}_M^{OS}\right]+C_M^{OS}\right)+\tilde{\alpha}_i\dfrac{(1+\beta)^2 Var\left[\tilde{C}_R^{OS}\right]}{2} & ,if\ OS\ is\ selected \\ P^{FS} & ,if\ FS\ is\ selected\end{cases} \quad (20)$$

where $\tilde{u}_i(B)$ is the $i's$ customer's disutility for OS and FS. The disutility is formulated by the sum of the expected price paid by clients and a penalty term for price volatility, the latter only applicable to OS. The penalty varies from one customer to another due to the heterogeneous individual risk aversion factor $\alpha_i$. $\beta$ is a mark-up factor in terms of accounting theory, assuming the OEM applying cost-plus pricing for OS service.

**Proposition 1** By the inferences above, OEM's profit is as follows:

$$\pi(P^{FS})\begin{cases}\beta\left(E\left[\tilde{C}_M^{OS}\right]+C_M^{OS}\right), \\ \tilde{\alpha}_i<\dfrac{2\left[p^{FS}-(1+\beta)\left[E\left[\tilde{C}_M^{OS}\right]+C_M^{OS}\right]\right]}{(1+\beta)^2 Var\left[\tilde{C}_R^{OS}\right]} \\ p^{FS}-E\left[\tilde{C}_R^{FS}(A(B),M^*)\right]-E\left[\tilde{C}_D^{FS}(M^*)\right]-\tilde{C}_M^{FS}(M^*)-C^l, \\ \tilde{\alpha}_i\geq\dfrac{2\left[P^{FS}-(1+\beta)\left[E\left[\tilde{C}_R^{OS}\right]+C_M^{OS}\right]\right]}{(1+\beta)^2 Var\left[\tilde{C}_R^{OS}\right]}\end{cases}$$

(21)

The first line in Equation (21) refers to OS profit, while the second line captures FS profit. The authors omit B in this equation as all costs refer to the entire contract time. The distinct superscripts OS and FS illustrate different expected repair and maintenance costs discussed above.

By referring to the study of Huber and Spinler (2014), the optimal full-service price $P^{FS*}$, based on the effects of learning and forgetting $A$, optimal maintenance policy $M^*$, and customer's risk attitudes $\alpha_i$ , is given by:

$$P^{FS^*} = \max\left\{ \min \left\{ \begin{array}{c} E\left[\tilde{C}_R^{FS}\left(A(B),M^*\right)\right] + \beta \cdot E\left[\tilde{C}_R^{OS}\right] + \frac{1}{2}C_M^{FS}\left(M^*\right) + \frac{1}{2}(1+2\beta)C_M^{OS} \\ + \frac{1}{2}E\left[\tilde{C}_D^{FS}\left(M^*\right)\right] + \frac{\alpha_{\max}(1+\beta)^2 Var\left[\tilde{C}_R^{OS}\right]}{4} + C^l, \overline{P^{FS}} \end{array} \right\}, \underline{P^{FS}} \right\}$$

(22)

**Proof**. See Appendix.

When the optimal solution is obtained as above, two observations can be made. First, FS prices are not only influenced by repair and maintenance costs led by FS, but also aggravated by the increasing of OS mark-up margin $\beta$. Second, the cost of delay in FS corresponds to the impact of cost variability that OS has on the customer with the greatest aversion. Both factors are stochastic as they measure how uncertainties or contingent issues in either FS or OS would affect the final price. Assuming that repair cost is reduced due to the learning or forgetting effect (A = 1), the non-optimized maintenance policy $M_0$ is applied for both OS and FS, and training cost is ignored, the price model is marked as $P_{bench}^{FS}$ which has been discussed previously in literature review. Combining $P_{bench}^{FS}$ with autonomous learning effect and optimized maintenance $M^*$ derives $P_{auto}^{FS}$ and $P^{FS}$ respectively, thus three FS prices exist.

After working out the optimal FS price, the lowest and highest limits deduced by Huber and Spinler (2014) are illustrated below:

$$P^{FS} \geq E\left[\tilde{C}_R^{FS}\left(A(B),M^*\right)\right] + E\left[\tilde{C}_D^{FS}\left(M^*\right)\right] + C_M^{FS}\left(M^*\right) + C^l + \beta\left(E\left[\tilde{C}_R^{OS}\right] + C_M^{OS}\right) \equiv \underline{P^{FS}} \quad (23)$$

$$\overline{P^{FS}} \equiv TCO - C_{Lease}(B) - C_{Ops}(B) \quad (24)$$

Obviously, a lower limit for the FS price $\underline{P}^{FS}$ follows from the fact that FS profit must be at least as large as the sum of OS profit and preventive, corrective, delay and training costs incurred by FS. A higher limit is given by a competitive benchmark in terms of total cost of ownership (TCO) for the device. TCO is composed of leasing costs $C_{Lease}(B)$, and operational costs $C_{Ops}(B)$, as well as maintenance and repair costs. Therefore, the higher limit FS price should be no greater than the TCO benchmark leasing and operational costs, otherwise clients are likely to discard the device.

***Proposition 2*** *Holding other factors fixed, because of training cost* $C^l$ *existence,* $P^{FS} < P^{FS}_{auto} < P^{FS}_{bench}$*, depends on training frequency* $lf$ *and unit training cost* $C^l$*. Because both two factors would directly affect the total training cost* $C^l$*, and once such cost becomes too large to be covered by efficiency improvements driven by induced learning, work place training could oppositely become a heavy burden to companies.*

**Proof**. See Appendix.

## 3. Numerical Analysis

### *3.1 Condition setting and data preparation*

This numerical study is based on working records obtained from an OEM that manufactures large medical devices, such as MRI and CT machines, mainly for tertiary hospitals. Firstly, this paper assumes that clients will choose either FS or OS and adhere to it for the whole lifetime of devices. According to GAAP (Generally Accepted Accounting Principle), the depreciation period is 10 years for those devices. This 10-year lifetime is divided into 20 equal segments. While the data provided are, for confidentiality reasons, not the true values, and some stochastic variables or parameters are simulated by computer programming, they are reasonably indicative of a realistic setting. The key parameters are set, shown in Table 2.

The following time-independent parameters are simulated and used: uniform distribution of risk aversion $\tilde{\alpha}_i \sim U\left[0,10^{-3}\right]$, OS price margin $\beta = 0.5$; the expected improvement factor for individual maintenance action $\rho = 0.5$; the external failure rate $\phi_{ext}$, which is simulated as a set of random numbers of normal distribution but with significantly lower values than the internal failure rate as it demonstrates accidental failures, autonomous and induced learning factors; $\alpha_{auto}$ and $\alpha_{indu}$, which are both 0.1, which equals the passive learning factors $\alpha$ in the existing model; and the learning exponent for reworks $\varepsilon = 0.05$, which is only half of the normal learning rate, as stated by Lolli et al. (2016). Besides the fixed values, the training cost coefficient (unit training cost) $l$ = \$50/hour. Time-independent individual costs include the unit maintenance cost $c_M$ = \$300.00, expected individual delay cost $E\left[\tilde{C}_d\right]$=\$10,000 and time-dependent parameter values, as provided in Table 2. The number of clients for both OS or FS together is set to $D = 50$, while the number of maintenance actions in OS is $M_0 = 10$ according to historical data from the OEMs who perform seasonal maintenance, and the beginning internal failure rate is $\Phi_0^{\text{int}} = 7.5 \times 10^{-3}$.

**Table 2.** Key parameters of the model

In this 20-period model, $t_j$ = 1440 hours and $t_j^M$ = 4320 hours are set for each period. This means that each period covers 180 days and the device runs 8 hours per day on average, with a 10-year contract for medical device maintenance provided by OEMs. Meanwhile, the internal and external failure rate $\phi^{\text{int}}$ and $\phi^{ext}$ are positive in all 20 periods, both of which are simulated. $\phi^{ext}$ is a set of random numbers and $\phi^{ext}$ is aligned with the bathtub distribution with different expected $E\left[\phi^{\text{int}}\right]$. The optimal number of maintenance $M^*$=3 times, and training frequency $lf$ simulated by a computer falls within [0.0031, 0.0250].

### *3.2 Numerical results and comparisons*

Initially set $E\left[\phi^{\text{int}}\right]$= 0.0034, $lf$ = 0.005. The key performance indicators (KPIs) including optimal price, cost and total profit for different pricing models, are listed and compared in Table 3. Three key consequences can be drawn from Table 3.

**Table 3.** KPIs of different pricing models

#### *3.2.1 General comparison between new and old FS model*

Compared with the old FS model solely considering autonomous learning effect, in the new one, the revenue to OEMs decreases by 14.5% while cost reduces by 23.1%, thus the profit brought by FS service finally increases 5.0%. In terms of market share, which is represented by 50% of clients with different risk appetite choosing full-time repair contracts, FS contracts with learning effects could attract a more significant number of clients switching from OS to FS than the benchmark. But for the market position, the introduction of induced learning and forgetting effects makes a slight difference, only 4%, in comparison with the existing model raised by Huber and Spinler (2012), though the price declines approximately 15%. As a result, the new FS model is more efficient in attracting clients and increasing profits than the old one.

### *3.2.2 Comparison between new, old FS model, and OS*

The insurance-like FS contract, without induced learning, forgetting or training costs, has a price premium to OS by 2%, because clients are willing to pay for the premium where cost variability of FS (zero) is much lower compared to the high OS cost variability. At the same time, the profit of FS in this old model becomes 13.8% higher than OS, but the FS cost is slightly lower at around 4.17%. When the new model adapted in this paper is applied, even with the training expense, the optimal repair price would be 12.8% lower than OS. A lower price than the old model due to higher efficiency from induced learning would obviously attract new clients switching from an OS to an FS contract. Interestingly, the adjusted FS model contributes to 14.3% higher profit than OS even though the price significantly decreases, and the cost to OEMs incurred by a FS contract surprisingly becomes 22.7% lower than OS. Consequently, the new FS model is more competitive in pricing than the old FS model and OS.

## ***3.3 Sensitivity analysis***

Training cost $C^l$ exists in the new model with induced learning and forgetting effects. By varying expected internal failure $E\left[\phi^{\text{int}}\right]$ and training frequency $lf$ defined in previous sections, the changes regarding the mark-up factor, the training frequency and unit training cost as well as their effect on FS revenue, FS cost and FS profit, become clear in Figures 1, 2, and 3 illustrated below.

### 3.3.1 Analysis on mark-up factor β

It is easy to understand that for the OS maintenance model, price and profit should be positively related to the price mark-up factor β, since the repair price in cost-plus approach determined by But after FS contract together with OS to the repair market are introduced, whether β still has the same effect on price, cost and profit becomes obscure so that requires further study. To address this, the author set learning rate *lr* = 0.005， and unit training cost *l* = 50 constantly; then pick 5 values ranging from 0.5 to 0.9, and another extra 5 values between 1.0 – 5.0.

**Table 4: KPI Results of sensitivity analysis to mark-up factor $\beta$ ($\beta \in$ [0.5 - 0.9])**

As the Table 3 show, if those several OEMs are in a competitive market, which suggests that price mark up factor only within a range of relatively value (set as [0.5,0.9] here), the optimal price and profit will be monotonically increasing with a constant slope. For cost, because $\beta$ is not involved as a parameter in the formula of cost, obviously, cost keep constant. For market position of FS contract, when $\beta$ belongs to [0.5,0.9], the FS market share seems not to be hurt by rising of $\beta$. In other word, OEMs could benefit more, for the burden of increased price mark-up factor lays on customers.

However, Table 5 below illustrates that if OEMs could form a oligopolistic market as price makers tacitly charge customers over twice as much as their repair cost (i.e. $\beta >$ 1), the market share of FS service, even though profit could keep rising, suffers a downward trend when $\beta$ falls within 1.0 - 3.0, until reaches a turning point 4.0 or more, after which OS becomes too expense for customers to bear even though FS is costly, too. The phenomenon indicates that in the real world of management, it could be not wise raise price overly high, as the market share of FS would be compromised and the turning point is usually unreachable in an competitive open market.

**Table 5: KPI Results of sensitivity analysis to mark-up factor $\beta$ ($\beta \in$ [1.0 – 5.0])**

*3.3.2 Analysis on internal failure rate $\phi^{\text{int}}$*

Price and cost are positively correlated to $\phi^{\text{int}}$. The times of corrective repair $E[\tilde{N}_j(M)]$ and optimal times of preventive maintenance $M^*$ respectively increase when the internal failure rate goes up in each stage, leading to the increase in both the cost of corrective repairs and preventive maintenance. Referring to Equation (24) in 2.4, the optimized FS price will increase until the maximum value of $P_{FS}$, whose detailed tendency is shown in Table 4 and Figure 1. To make this table and figure readable, expected value of each set of internal failure rates is set as $E\left[\phi^{\text{int}}\right]$, and each series of $\phi^{\text{int}}$ is provided in the appendix.

**Table 6. KPI Results of sensitivity analysis to Internal failure $\phi^{\text{int}}$**

Table 5 clearly depicts that more clients are driven to switch to FS repairs as the expected failure rate $E\left[\phi^{\text{int}}\right]$ gradually rises with a low initial value; meanwhile, price, cost and total profit increase with the market share of an FS contract. A rise in internal failure rate would lead to higher prices, costs and profits, even if market share peaks at 100% and remains steady. As is shown in Table 4 and Figure 1, optimal price illustrates a positive relationship with $\phi^{\text{int}}$. While the overall failure rate increases, according to Equation (20) in 2.4, the disutility of OS repairs goes up significantly because of volatility factor $\alpha_i \frac{(1+\beta)^2 Var[\tilde{C}_R^{OS}]}{2}$.

**Figure 1.** KPI Results of sensitivity analysis to Internal failure rate $\Phi^{\text{int}}$

Furthermore, higher disutility from OS repair drives more clients, especially those with a greater risk aversion factor $\alpha$, to switch from OS repairs to FS repairs, to transfers risk of volatility to the OEM. Naturally，they would accept to pay more than service utility for a FS contract, just like an insurance premium. This contributes higher profit margin to the OEM , as well as a higher market share of FS repair contracts. Futhermore, The increase in internal failure rate can significantly harm OS clients financially because the repair expenditure is completely borne by themselves, therefore, they are usually willing to pay a much higher premium to lock their cost in advance.

*3.3.3 Analysis on training frequency $lf$*

If only the training frequency $lf$ is changed and set $C^l = 50$, $E\left[\phi^{\text{int}}\right] = 0.0034$, we assume that both revenue and cost for FS increase after a brief downward trend. In contrast, the profit for FS remains constant disregarding the changes of learning rate.

As Table 6 and Figure 2 state, the optimal training frequency $lf$ falls within the interval of [0.0031, 0.0250], among which the new model with induced learning and forgetting effects in this paper gains the greatest advantage over the original model of Huber and Spinler (2012). However, the rise in learning rate makes neither positive or negative contributions to the total profit.

**Table 7.** KPIs Result of sensitivity analysis to training frequency $lf$

**Figure 2.** KPIs Result of sensitivity analysis to training frequency $lf$

In addition, according to Table 6 and Figure 2, when the cumulative time of training remains at a low level, the increase in cumulative training time can provide more attractive benefits to both the customer, through a lower price, and the companies, by a lower cost and greater profit. However, once the amount of total training time exceeds the turning point, the cost and price would soar while profit would decline sharply.

Therefore, the induced learning would totally lose its competitiveness if training time is overly long. An OEM might lose its profits due to lengthy training time and high training cost, and eventually lose its core competitiveness, which validate Huber and Spinler's (2014) statement

**Figure 3.** Profit premium over previous model unit training cost =$100/h

**4. Significances and Conclusion :**

This paper improves the existing FS pricing model, which Huber & Spiner (2012, 2014) initially founded based only on simple autonomous learning , by introducing induced learning process and dynamic forgetting effects.

The numerical analysis shows that with the optimal price decreasing, cost reduces more significantly so that the total profit increases when our new FS model is applied, in comparison with the old model with solely autonomous learning. Supposed that 50% clients choose full-time repair contracts in market, the introduction of induced learning and forgetting effects makes no difference compared with the old model, though the price declines. Additionally, the results also illustrate the advantages of the enhanced FS model by comparing it with previous autonomous learning-only FS model, and OS respectively. At the very beginning, clients who signed the FS contracts are ready to accept a higher warranty premium because they take lower risks since the variability of maintenance cost of FS contract is lower than its OS counterparts. Moreover, even training expenses exists, the optimal price of FS maintanence contract is lower than that of OS thanks to the improved learning efficiency through scheduled traning. Both of the two aspects attract can stimulate new clients to switch from an OS maintenance to an FS contracts. And, as is shown on sensitivity analysis, the optimal price reflects a positive correlation with $\phi^{\text{int}}$

. When the optimal training frequency $lf$ is applied, the new FS model discussed in this study take the greatest advantage over both the original FS and OS models .

In addition to the theoretical and numerical contributions, the improved FS contract shall **also benefit the real world of management**. After introducing the external training procedures, technicians of OEMs can not only learn by performing daily duties, but also by receiving external vocational training (aka. induced learning) provided by senior staff or outsource labour agencies. Nevertheless, the true effectiveness of autonomous learning is usually unpredictable or uncontrollable; the training investment and procedure, instead, can be well monitored by management. Supervisor may track the effectiveness of external training through routine performance reviews, vocational exams, even contests.

In summary, the study shows that application of the improved FS pricing model, which consists of autonomous, induced learning as well as forgetting effects, plays a prominent role in OEMs' management decision making towards providing maintenance service to clients. Ultimately, rmore optimal price , lower cost , higher profit and greater market share are expected .

**Acknowledgements**

The research described in this paper has been funded by the National Natural Science Foundation of China (Grant No. 71302053).

**Disclosure statement**

No potential conflict of interest was reported by the authors.

## Appendix

**Proof of Proposition 1:**

It follows from Equation (22), with $\tilde{\alpha}_i \sim U\left[0, \alpha_{maz}\right]$ and D representing the market size, that the OEMs maximize their service profit:

$$\underset{P^{FS}\geq 0}{\operatorname{argmax}} - E\left[\pi\left(P^{FS}\right)\right] = D\Big\{ \left[P^{FS} - E\left[\tilde{C}_R^{FS}\left(A(B), M^*\right)\right] - E\left[\tilde{C}_D^{FS}\left(A(B), M^*\right)\right] - \tilde{C}_M^{FS}\left(M^*\right) - C^l\right] \times \Pr\left(\tilde{\alpha}_i \geq \frac{2\left[P^{FS} - (1+\beta)\left[E\left[\tilde{C}_R^{OS}\right] + C_M^{OS}\right]\right]}{(1+\beta)^2 Var\left[\tilde{C}_R^{OS}\right]}\right)$$

$$+\beta\left(E\left[\tilde{C}_R^{OS}\right] + C_M^{OS}\right) \times \Pr\left(\tilde{\alpha}_i \leq \frac{2\left[P^{FS} - (1+\beta)\left[E\left[\tilde{C}_R^{OS}\right] + C_M^{OS}\right]\right]}{(1+\beta)^2 Var\left[\tilde{C}_R^{OS}\right]}\right)\Big\}$$

$$= D\Big\{ \left[P^{FS} - E\left[\tilde{C}_R^{FS}\left(A(B), M^*\right)\right] - E\left[\tilde{C}_D^{FS}\left(A(B), M^*\right)\right] - \tilde{C}_M^{FS}\left(M^*\right) - C^l\right] \times \left(1 - \frac{2\left[P^{FS} - (1+\beta)\left[E\left[\tilde{C}_R^{OS}\right] + C_M^{OS}\right]\right]}{\alpha_{\max}(1+\beta)^2 Var\left[\tilde{C}_R^{OS}\right]}\right)$$

$$+\beta\left(E\left[\tilde{C}_R^{OS}\right] + C_M^{OS}\right) \times \left(\frac{2\left[P^{FS} - (1+\beta)\left[E\left[\tilde{C}_R^{OS}\right] + C_M^{OS}\right]\right]}{\alpha_{\max}(1+\beta)^2 Var\left[\tilde{C}_R^{OS}\right]}\right)\Big\}$$

s.t. $\underline{P}^{FS} \leq P_{auto}^{FS} \leq P_{bench}^{FS}$

Omitting the factor D, the function above yields the following first-order derivative:

$$\frac{\partial}{\partial P^{FS}} = 1 - \frac{2\left[P^{FS} - (1+\beta)\right]\left[E\left[\tilde{C}_R^{OS}\right] + C_M^{OS}\right] + 2\left[P^{FS} - E\left[\tilde{C}_R^{FS}\left(A(B), M^*\right)\right] - E\left[\tilde{C}_D^{FS}\left(M^*\right)\right] - \tilde{C}_M^{FS}\left(M^*\right) - C^l\right]}{\alpha_{\max}(1+\beta)^2 Var\left[\tilde{C}_R^{OS}\right]} \pm \frac{2\beta\left(E\left[\tilde{C}_R^{OS}\right] C_M^{OS}\right)}{\alpha_{\max}(1+\beta)^2 Var\left[\tilde{C}_R^{OS}\right]}$$

$$= 1 - \frac{4P^{FS} - 2(1+\beta)\left[E\left[\tilde{C}_R^{OS}\right] + C_M^{OS}\right] - 2\left[E\left[\tilde{C}_R^{FS}\left(A(B), M^*\right)\right] + E\left[\tilde{C}_D^{FS}\left(M^*\right)\right] + \tilde{C}_M^{FS}\left(M^*\right) + C^l\right]}{\alpha_{\max}(1+\beta)^2 Var\left[\tilde{C}_R^{OS}\right]} + \frac{2\beta\left(E\left[\tilde{C}_R^{OS}\right] C_M^{OS}\right)}{\alpha_{\max}(1+\beta)^2 Var\left[\tilde{C}_R^{OS}\right]}$$

$$= 1 + \frac{2(1+\beta)\left[E\left[\tilde{C}_R^{OS}\right] + C_M^{OS}\right] + 2\left[E\left[\tilde{C}_R^{FS}\left(A(B), M^*\right)\right] + E\left[\tilde{C}_D^{FS}\left(M^*\right)\right] + \tilde{C}_M^{FS}\left(M^*\right) + C^l\right]}{\alpha_{\max}(1+\beta)^2 Var\left[\tilde{C}_R^{OS}\right]} - \frac{4P^{FS}}{\alpha_{\max}(1+\beta)^2 Var\left[\tilde{C}_R^{OS}\right]} + \frac{2\beta\left(E\left[\tilde{C}_R^{OS}\right] C_M^{OS}\right)}{\alpha_{\max}(1+\beta)^2 Var\left[\tilde{C}_R^{OS}\right]} \overset{set}{=} 0$$

The solution is unique as the maximization problem is quadratic. Prove up.

**Proof of Proposition 2** First of all, we need to prove that $P_{auto}^{FS} \leq P_{bench}^{FS}$ exists. To achieve this, we need to show:

$$(1+\beta)\left(E\left[\tilde{C}_R^{bench}\right]+E\left[\tilde{C}_M^{OS}\right]\right)+E\left[\tilde{C}_D\right]\geq E\left[\tilde{C}_R^{FS}\left(A_{auto}(B),M^*\right)\right]+E\left[\tilde{C}_D^{FS}\left(M^*\right)\right]$$
$$+\tilde{C}_M^{FS}\left(M^*\right)+\beta\left(E\left[\tilde{C}_R^{OS}\right]C_M^{OS}\right)$$

. Obviously, there exist the relationship: due to autonomous learning effect marked as $A_{auto}$ which lowers the repair cost. Moreover, assuming that $A=1$ and $M=M_0$, $E\left[\tilde{C}_R^{bench}\right]=E\left[\tilde{C}_R^{OS}\right]$.

Then the problem becomes to show whether and how $\underline{P^{FS}}\leq\underline{P_{auto}^{FS}}$ could be held. As it is assumed that there is training cost $C^l$,the in-equation in proof 2 need to be adjusted as

$$E\left[\tilde{C}_R^{FS}\left(A_{auto}(B),M^*\right)\right]+E\left[\tilde{C}_D^{FS}\left(M^*\right)\right]+\tilde{C}_M^{FS}\left(M^*\right)+\beta\left(E\left[\tilde{C}_R^{OS}\right]C_M^{OS}\right)\geq E\left[\tilde{C}_R^{FS}\left(A(B),M^*\right)\right]+E\left[\tilde{C}_D^{FS}\left(M^*\right)\right]+\tilde{C}_M^{FS}\left(M^*\right)+C^j+\beta\left(E\left[\tilde{C}_R^{OS}\right]C_M^{OS}\right)$$

.Whether this inequation can be held solely depends on the value of training cost $C^l$ .Moreover, from the inference in section 3.1.2, $C^l$ is positively correlated to Training frequency $lf$ and unit training cost $l$. Once either of those two parameters exceed some break-even point, $C^l$ will become too large so that the direction of in-equation will invert, then $\underline{P^{FS}}\leq\underline{P_{auto}^{FS}}$ bench fails.

Table 6 demonstrates 10 different series of $\phi^{\text{int}}$ which simulates internal failures in heterogeneous situations. Each series is given as value of expectation in the main body of article.

Word count: 8205 words

Table 1. Notation used in the model

| Random variables | Descriptions |
|---|---|
| $C_R$ | Stochastic repair costs over contract period |
| $C_{rj}$ | Individual stochastic repair costs over the $j$'s maintenance period |
| $N_j$ | Number of failures in the $j$'s period |
| $N$ | Number of failures over contract period |
| $\Phi_j^{\text{int}}$ | Likelihood of internal failure in the $j$'s period |
| $\Phi_j^{ext}$ | Likelihood of external failure in the $j$'s period |
| $C_D$ | Stochastic delay costs over contract period |
| $C_d$ | Individual stochastic delay costs over the whole contract period |
| **Decision variable** | **Description** |
| $P^{FS}$ | FS price |
| **Other Variables** | **Descriptions** |
| $j$ | Period index ( $j = 1, 2, 3...$) |
| $Z$ | The number of periods |
| $z_i$ | Stage index ( $i = 1, 2, 3, ...$) |
| $g_j$ | Ageing factor in the $j$'s period |
| $M(M^*)$ | Number of maintenance activities for FS (optimized) |
| $A$ | Cumulative learning effect over equipment contract period |
| $B$ | Length of the whole contract period |
| $N(t)$ | Cumulative units of natural learning before and including time $t$ |
| $L(t)$ | Cumulative units of induced learning before and including time $t$ |
| $t^r$ | Total length of repair time |
| $t^l$ | Total length of training time |
| $t^f$ | Total length of forgetting time |
| $lf$ | Learning rate factor |
| $C_{Lease}$ | Leasing costs |
| $C_{Ops}$ | Operational costs |
| $\pi$ | OEM's profit |

| Parameters | Descriptions |
|---|---|
| $\varepsilon$ | Learning factor for reworks |
| $\alpha$ | Learning factor |
| $\rho$ | Maintenance improvement factor |
| $t_j$ | Length of the $j$'s period |
| $t_j^M$ | Length of the $j$'s maintenance period |
| $\bar{c}_m$ | Average cost of single maintenance |
| Symbol | Description |
| $E[\cdot]$ | Expectation operator |

Table 2. Key parameters of the model

| $\beta$ | $\rho$ | $\alpha$ | $\alpha_{auto}$ | $\alpha_{indu}$ | $\varepsilon$ |
|---|---|---|---|---|---|
| 0.5 | 0.5 | 0.1 | 0.1 | 0.1 | 0.05 |
| $l$ ($/hour) | $c_M$ ($) | $E[\tilde{C}_d]$ | $D$ | $M_0$ | $\Phi_0^{\text{int}}$ |
| 50 | 300 | 10000 | 50 | 10 | $7.5\times10^{-3}$ |

Table 3. KPIs of different pricing models

| Pricing Model / KPIs (In thousand $) | New Pricing Model | Old Pricing Model | On-Call Service Model |
|---|---|---|---|
| **Optimal Price $P^{FS*}$** | 198 | 231 | 227 |
| **Cost C** | 111 | 145 | 151 |
| **Total Profit, D=50** | 4323 | 4302 | 3781 |
| **FS market share** | 100% | 96% | NA |

**Table 4: KPI Results of sensitivity analysis to mark-up factor $\beta$ ($\beta \in$ [0.5 - 0.9])**

| *KPI ($000)* / *Value of $\beta$* | *Optimal Price* $P^{FS*}$ | *Cost* $C^{FS*}$ | *Total Profit* $\pi(P^{FS})^*$ | *FS Market Share* |
|---|---|---|---|---|
| 0.5 | 198 | 111. | 4323 | 100% |
| 0.6 | 217 | 111 | 5277 | 100% |
| 0.7 | 237 | 111 | 6243 | 100% |
| 0.8 | 256 | 111 | 7222 | 100% |
| 0.9 | 276 | 111 | 8213 | 100% |

**Table 5: KPI Results of sensitivity analysis to mark-up factor $\beta$ ($\beta \in$ [1.0 – 5.0])**

| *KPI (($000)* / *Value of $\beta$* | *Optimal Price* $P^{FS*}$ | *Cost* $C^{FS*}$ | *Total Profit* $\pi(P^{FS})^*$ | *FS Market Share* |
|---|---|---|---|---|
| 1.0 | 296 | 111.69 | 9218 | 96% |
| 2.0 | 511 | 111 | 19187 | 78% |
| 3.0 | 751 | 111 | 29560 | 70% |
| 4.0 | 900 | 111 | 38130 | 86% |
| 5.0 | 900 | 111 | 39438 | 100% |

Table 6. KPI Results of sensitivity analysis to Internal failure $\phi^{\text{int}}$

| *KPI($000 )* | *Optimal* | *Cost* | *Total* | **FS Market** |
|---|---|---|---|---|

| $E\left[\phi^{\text{int}}\right]$ | ***Price*** $P^{FS*}$ | $C^{FS*}$ | ***Profit*** $\pi(P^{FS})^*$ | **Share** |
|---|---|---|---|---|
| 0.0019 | 127 | 85 | 2014 | 52% |
| 0.0022 | 137 | 86 | 2455 | 68% |
| 0.0025 | 157 | 104 | 2529 | 46% |
| 0.0028 | 183 | 109 | 3704 | 100% |
| 0.0031 | 189 | 112 | 3876 | 100% |
| 0.0034 | 198 | 112 | 4323 | 100% |
| 0.0037 | 220 | 130 | 4492 | 100% |
| 0.0040 | 253 | 135 | 5913 | 100% |
| 0.0043 | 258 | 136 | 6093 | 100% |
| 0.0046 | 282 | 151 | 6522 | 100% |

Table 7. KPIs Result of sensitivity analysis to training frequency $lf$

| KPI($000) / Value of $lf$ | Optimal Price $P^{FS*}$ | Cost $C^{FS*}$ | Total Profit $\pi(P^{FS})^*$ | FS Market Share |
|---|---|---|---|---|
| 0.0031 | 199.3 | 112.9 | 4323 | 100% |
| 0.0033 | 198.9 | 112.4 | 4323 | 100% |
| 0.0036 | 199.0 | 112.6 | 4323 | 100% |
| 0.0038 | 198.4 | 112.0 | 4323 | 100% |
| 0.0042 | 198.1 | 112.7 | 4323 | 100% |
| 0.0045 | 198.1 | 111.6 | 4323 | 100% |
| 0.0050 | 198.1 | 111.6 | 4323 | 100% |
| 0.0056 | 197.7 | 111.2 | 4323 | 100% |
| 0.0063 | 198.4 | 111.9 | 4323 | 100% |
| 0.0071 | 198.2 | 111.7 | 4323 | 100% |
| 0.0083 | 198.9 | 112.4 | 4323 | 100% |
| 0.0100 | 200.1 | 113.6 | 4323 | 100% |
| 0.0125 | 202.4 | 115.9 | 4323 | 96% |
| 0.0167 | 206.4 | 120.0 | 4323 | 80% |
| 0.0250 | 216.1 | 129.6 | 4323 | 32% |

Table 8. 10 series of $\phi^{int}$

| $\phi_1^{int}$ | $\phi_2^{int}$ | $\phi_3^{int}$ | $\phi_4^{int}$ | $\phi_5^{int}$ | $\phi_6^{int}$ | $\phi_7^{int}$ | $\phi_8^{int}$ | $\phi_9^{int}$ | $\phi_{10}^{int}$ |
|---|---|---|---|---|---|---|---|---|---|
| 0.0046 | 0.0049 | 0.0052 | 0.0055 | 0.0058 | 0.0061 | 0.0064 | 0.0067 | 0.0070 | 0.0073 |
| 0.0034 | 0.0037 | 0.0040 | 0.0043 | 0.0046 | 0.0049 | 0.0052 | 0.0055 | 0.0058 | 0.0061 |
| 0.0022 | 0.0025 | 0.0028 | 0.0034 | 0.0034 | 0.0037 | 0.0040 | 0.0043 | 0.0046 | 0.0049 |
| 0.0011 | 0.0014 | 0.0017 | 0.0020 | 0.0023 | 0.0015 | 0.0029 | 0.0032 | 0.0035 | 0.0038 |
| 2.99e-05 | 0.0003 | 0.0006 | 0.0009 | 0.0012 | 0.0015 | 0.0018 | 0.0021 | 0.0024 | 0.0027 |
| 2.99e-05 | 0.0003 | 0.0006 | 0.0009 | 0.0012 | 0.0015 | 0.0018 | 0.0021 | 0.0024 | 0.0027 |
| 2.99e-05 | 0.0003 | 0.0006 | 0.0009 | 0.0012 | 0.0015 | 0.0018 | 0.0021 | 0.0024 | 0.0027 |
| 2.99e-05 | 0.0003 | 0.0006 | 0.0009 | 0.0012 | 0.0015 | 0.0018 | 0.0021 | 0.0024 | 0.0027 |
| 2.99e-05 | 0.0003 | 0.0006 | 0.0009 | 0.0012 | 0.0015 | 0.0018 | 0.0021 | 0.0024 | 0.0027 |
| 2.99e-05 | 0.0003 | 0.0006 | 0.0009 | 0.0012 | 0.0015 | 0.0018 | 0.0021 | 0.0024 | 0.0027 |
| 2.99e-05 | 0.0003 | 0.0006 | 0.0009 | 0.0012 | 0.0015 | 0.0018 | 0.0021 | 0.0024 | 0.0027 |
| 2.99e-05 | 0.0003 | 0.0006 | 0.0009 | 0.0012 | 0.0015 | 0.0018 | 0.0021 | 0.0024 | 0.0027 |
| 2.99e-05 | 0.0003 | 0.0006 | 0.0009 | 0.0012 | 0.0015 | 0.0018 | 0.0021 | 0.0024 | 0.0027 |
| 2.99e-05 | 0.0003 | 0.0006 | 0.0009 | 0.0012 | 0.0015 | 0.0018 | 0.0021 | 0.0024 | 0.0027 |
| 2.99e-05 | 0.0003 | 0.0006 | 0.0009 | 0.0012 | 0.0015 | 0.0018 | 0.0021 | 0.0024 | 0.0027 |
| 0.0016 | 0.0019 | 0.0022 | 0.0025 | 0.0028 | 0.0031 | 0.0034 | 0.0037 | 0.0040 | 0.0043 |
| 0.0033 | 0.0036 | 0.0039 | 0.0042 | 0.0045 | 0.0048 | 0.0051 | 0.0054 | 0.0057 | 0.0060 |
| 0.0051 | 0.0054 | 0.0057 | 0.0060 | 0.0063 | 0.0066 | 0.0069 | 0.0072 | 0.0075 | 0.0078 |
| 0.0070 | 0.0073 | 0.0076 | 0.0079 | 0.0082 | 0.0085 | 0.0088 | 0.0091 | 0.0094 | 0.0097 |
| 0.0091 | 0.0094 | 0.0097 | 0.0100 | 0.0103 | 0.0106 | 0.0109 | 0.0112 | 0.0115 | 0.0118 |

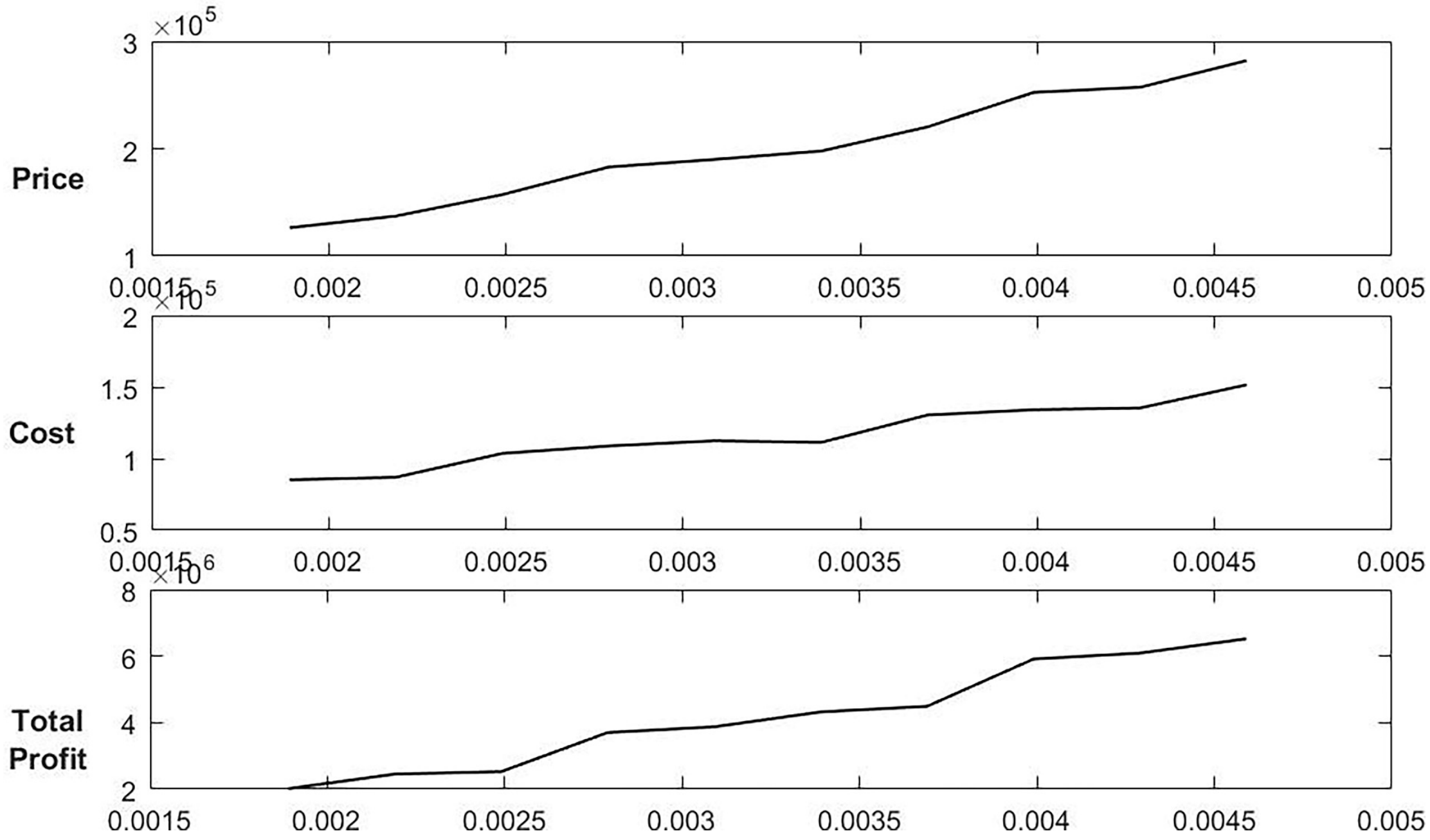

Figure 1. KPI Results of sensitivity analysis to Internal failure rate $\Phi^{int}$

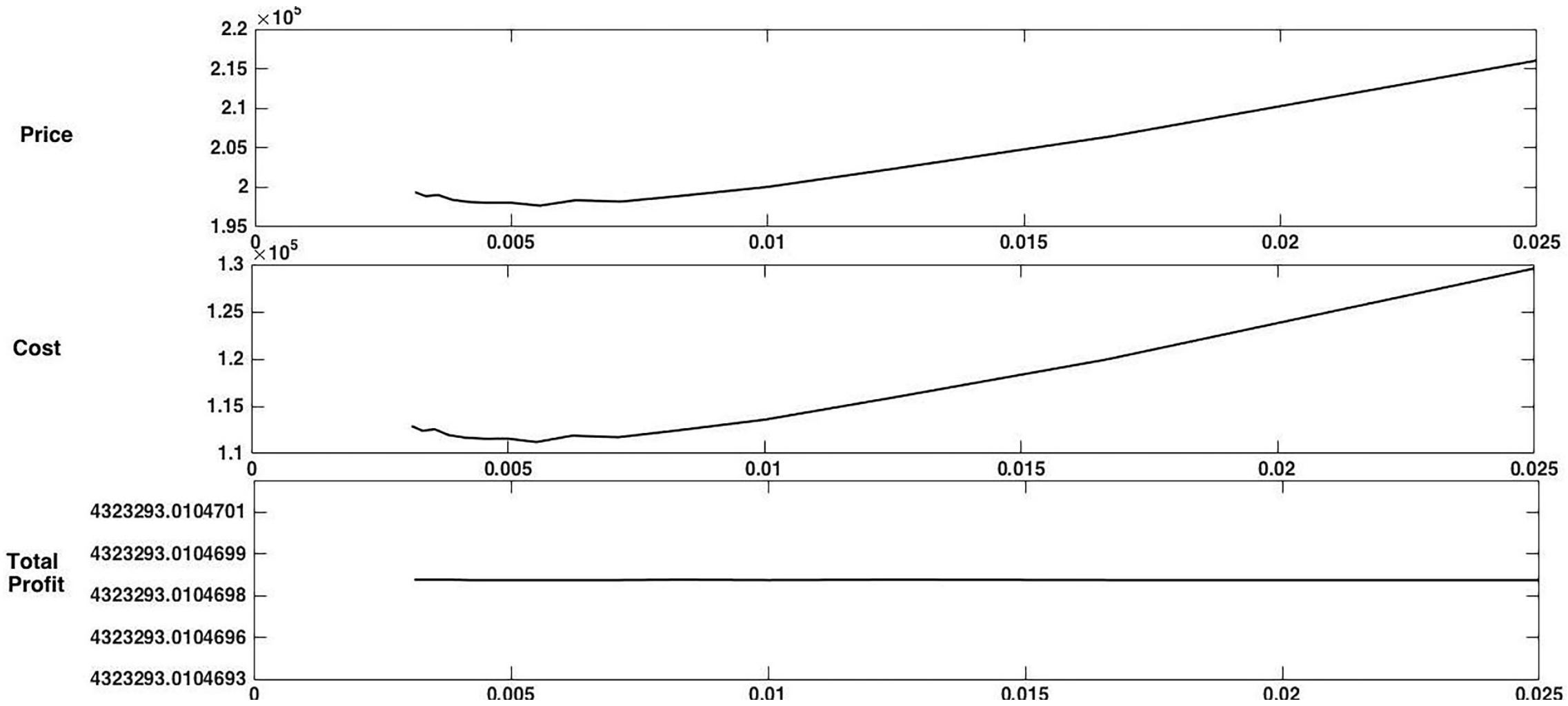

Figure 2. KPIs Result of sensitivity analysis to training frequency $lf$

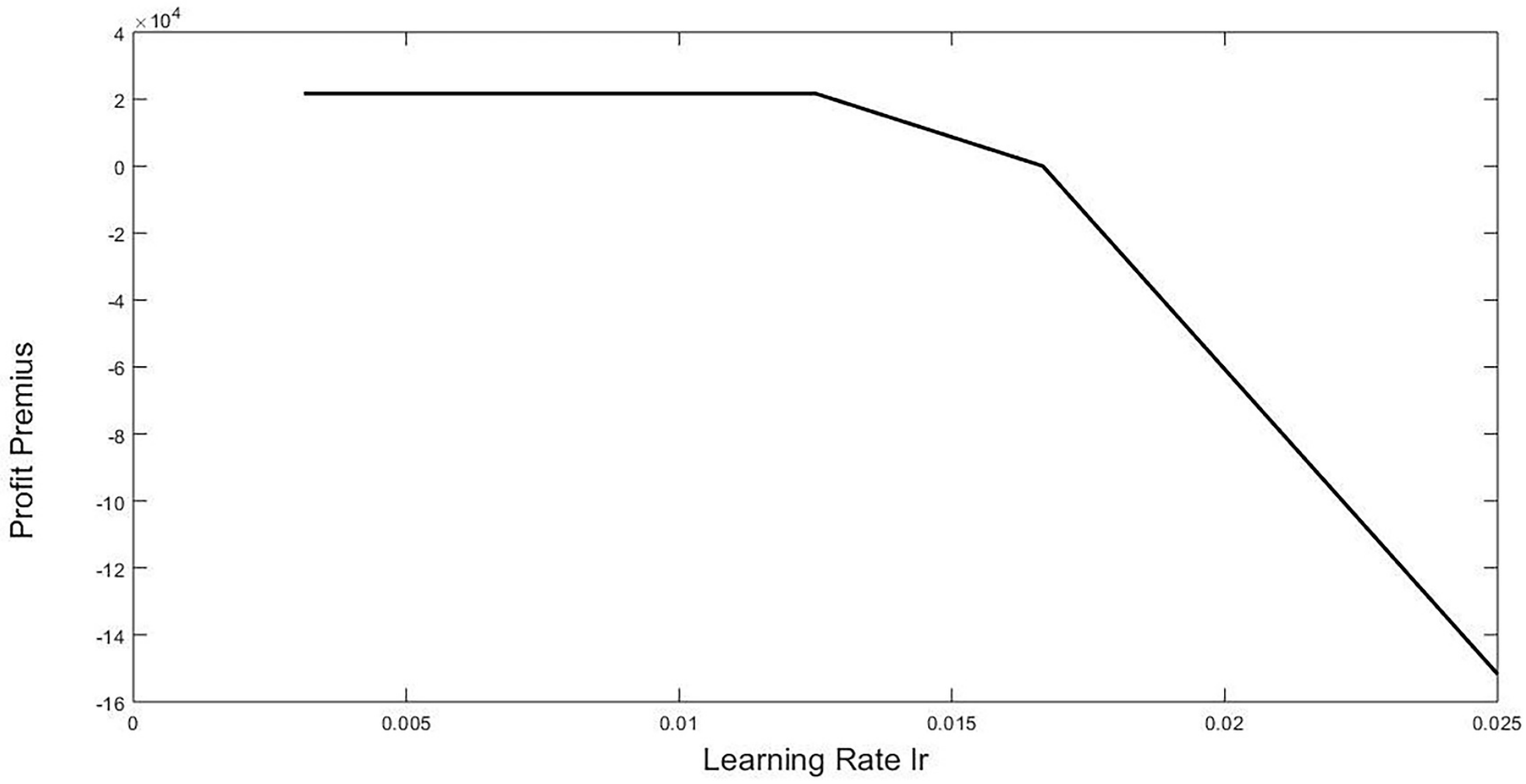

Figure 3. Profit premium over previous model unit training cost=100$/hr